\documentclass[conference]{IEEEtran}

\usepackage[table]{xcolor}
\usepackage{tikz}
\usepackage{lipsum} 
\usepackage{multicol} 
\usetikzlibrary{shapes, arrows, positioning}
\usetikzlibrary{patterns}

\usepackage{breqn}
\usepackage{bm}

\usepackage{mathtools}
\usepackage[T1]{fontenc}
\usepackage{comment}
\usepackage{threeparttable}

\usepackage{pgfplots}
\usepackage{pgfplotstable}

\usepackage{booktabs}
\usepackage{caption}

\pgfplotsset{compat=1.16}

\usepackage{float}
\usepackage{xcolor}
\usepackage{graphicx}
\usepackage{amssymb,stmaryrd}
\usepackage{halloweenmath}
\usepackage{subcaption}
\usepackage{amsthm}
\usepackage{amsfonts}
\usepackage[ruled,linesnumbered]{algorithm2e}

\usepackage[ruled,linesnumbered]{algorithm2e}

\let\oldnl\nl
\newcommand{\nonl}{\renewcommand{\nl}{\let\nl\oldnl}}
\SetKwInput{KwData}{Input}
\SetKwInput{KwResult}{Output}
\SetKwComment{Comment}{/* }{ */}
\DontPrintSemicolon

\SetCommentSty{mycommfont}

\usepackage{xcolor}
\usepackage{url}
\usepackage{booktabs}
\newtheorem{proposition}{Proposition}

\usepackage{mathtools}

\definecolor{codegreen}{rgb}{0,0.6,0}
\definecolor{codegray}{rgb}{0.5,0.5,0.5}
\definecolor{codepurple}{rgb}{0.58,0,0.82}
\definecolor{backcolour}{rgb}{0.95,0.95,0.92}
\usepackage{listings}
\lstdefinestyle{mystyle}{
    backgroundcolor=\color{backcolour},   
    commentstyle=\color{codegreen},
    keywordstyle=\color{magenta},
    numberstyle=\tiny\color{codegray},
    stringstyle=\color{codepurple},
    basicstyle=\ttfamily\footnotesize,
    breakatwhitespace=false,         
    breaklines=true,                 
    captionpos=b,                    
    keepspaces=true,                 
    numbers=left,                    
    numbersep=5pt,                  
    showspaces=false,                
    showstringspaces=false,
    showtabs=false,                  
    tabsize=2
}

\theoremstyle{definition}

\usepackage[utf8]{inputenc}

\usepackage{seqsplit}

\usepackage{latexsym}

\usepackage{tikz}

\begin{document}
%
\title{Silca: Singular Caching of Homomorphic Encryption for Outsourced Databases in Cloud Computing}


\author{\IEEEauthorblockN{Dongfang Zhao}
\IEEEauthorblockA{University of Washington\\
dzhao@uw.edu}

}

\IEEEoverridecommandlockouts
\makeatletter\def\@IEEEpubidpullup{6.5\baselineskip}\makeatother
\IEEEpubid{\parbox{\columnwidth}{
    Network and Distributed System Security (NDSS) Symposium 2024\\
    26 February - 1 March 2024, San Diego, CA, USA\\
    ISBN 1-891562-93-2\\
    https://dx.doi.org/10.14722/ndss.2024.23xxx\\
    www.ndss-symposium.org
}
\hspace{\columnsep}\makebox[\columnwidth]{}}

\maketitle

\begin{abstract}
Ensuring the confidentiality and privacy of sensitive information in cloud computing and outsourced databases is crucial. Homomorphic encryption (HE) offers a solution by enabling computations on encrypted data without decryption, allowing secure outsourcing while maintaining data confidentiality. However, HE faces performance challenges in query-intensive databases. To address this, we propose two novel optimizations, Silca and SilcaZ, tailored to outsourced databases in cloud computing. Silca utilizes a singular caching technique to reduce computational overhead, while SilcaZ leverages modular arithmetic operations to ensure the applicability of singular caching for intensive HE operations. We prove the semantic security of Silca and SilcaZ and implement them with CKKS and BGV in HElib as MySQL loadable functions. Extensive experiments with seven real-world datasets demonstrate their superior performance compared to existing HE schemes, bridging the gap between theoretical advancements and practical applications in applying HE schemes on outsourced databases in cloud computing.
\end{abstract}

\section{Introduction}

With the increasing reliance on cloud computing and the outsourcing of data storage and processing~\cite{hhaci_icde02}, ensuring the confidentiality and privacy of sensitive information has become a critical concern~\cite{popa2011cryptdb},
particularly for applications that handle sensitive data in fields such as public health~\cite{tkanwal_cluster21}, bioinformatics~\cite{xzhu_tdsc21}, and financial services~\cite{wkuan_clsr18}. 
Homomorphic encryption (HE)~\cite{cgentry_stoc09}, a revolutionary cryptographic technique, offers a promising solution by enabling computations on encrypted data without the need for decryption. This unique capability allows data owners to securely outsource their databases to untrusted cloud providers while maintaining data confidentiality when conducting meaningful algebraic operations directly on the ciphertexts stored on the outsourced databases~\cite{symmetria_vldb20}.

Despite its significant potential, homomorphic encryption faces performance challenges, particularly in the context of cloud-based query-intensive outsourced databases~\cite{otawose_sigmod23}. The computational overhead incurred during encryption, decryption, and computation operations limits the practicality and scalability of HE schemes in real-world scenarios. Therefore, optimizing the encryption performance of HE schemes has become a crucial research direction.

In this work, our focus lies on improving the encryption performance of homomorphic encryption specifically in the context of cloud-based outsourced databases. We propose two novel optimizations, Silca (singular caching) and SilcaZ (singular caching for integers), aimed at addressing the performance limitations and enhancing the efficiency of HE. These optimizations leverage innovative techniques and algorithms tailored to the unique requirements and characteristics of outsourced databases in cloud computing.

The Silca optimization employs advanced caching strategies and indexing techniques to reduce the computational overhead associated with homomorphic encryption operations. 
Specifically, Silca observes that an arbitrary floating-point number can be decomposed into two factors,
one of which could a random value whose ciphertext can be precomputed in an offline stage.
Such multiplicative property can be carried to the ciphertext space as well if the underlying encryption scheme is homomorphic on the multiplication operation,
which is indeed the case for modern HE schemes,
such as CKKS~\cite{ckks}.
By intelligently managing the encrypted data and optimizing the access patterns, Silca minimizes the latency and improves the overall performance of computations on encrypted data.

Building upon the Silca optimization, we further introduce the SilcaZ optimization, which takes advantage of efficient modular arithmetic operations on group $\mathbb{Z}_p$ ($p$ denotes a prime number) and probabilistic data structures to enhance the encryption performance of homomorphic encryption in the context of query-intensive outsourced databases. 
SilcaZ is particularly useful when the applications require exact computation rather than approximate computation;
a good example of such a HE scheme is BGV~\cite{bgv}.
SilcaZ introduces efficient query processing techniques that minimize the communication and computational costs while ensuring the privacy and integrity of the outsourced data.

We demonstrate that Silca (and its varient SilcaZ) is semantic secure by reducing a presumably secure base HE scheme to Silca.
That is, we assume that an adversary $\mathcal{A}$ exists to break a Silca instance;
then, we show that $\mathcal{A}$ can also break the security of the corresponding base HE scheme,
which leads to a contradiction.

We implement both Silca and SilcaZ with HElib~\cite{helib} as loadable functions in MySQL 8.0.
Our extensive evaluation demonstrates the significant performance improvements achieved by Silca and SilcaZ compared to existing state-of-the-art schemes. In particular, Silca and SilcaZ surpass the performance of the state-of-the-art scheme Rache~\cite{otawose_sigmod23} by a factor of up to 10 times. This substantial enhancement showcases the effectiveness of the proposed optimizations in reducing computational overhead and improving efficiency.
Furthermore, Silca and SilcaZ also outperform the original fully homomorphic encryption (FHE) schemes, such as CKKS~\cite{ckks} and BGV~\cite{bgv}, by orders of magnitude. The execution time and resource utilization achieved by Silca and SilcaZ far exceed those of traditional FHE schemes, making them highly practical and scalable for real-world scenarios involving large-scale outsourced databases.

These remarkable results highlight the significant contribution of Silca and SilcaZ in bridging the gap between theoretical advancements in homomorphic encryption and their practical applications in the context of outsourced databases. The optimizations provided by Silca and SilcaZ offer a compelling solution for improving the encryption performance of homomorphic encryption techniques, making them highly promising for secure data outsourcing in cloud computing environments.

In summary, this paper makes the following contributions.
\begin{itemize}
    \item We propose two new caching algorithms, namely Silca and SilcaZ, to optimize the performance of homomorphic encryption in outsourced databases on the cloud.

    \item We prove the semantic security of both Silca and SilcaZ by reducing the base HE scheme to the proposed encryption schemes.

    \item We implement both Silca and SilcaZ with HElib as loadable functions in MySQL and demonstrate their effectiveness with seven datasets.
\end{itemize}

The remainder of this paper is organized as follows. In Section~\S\ref{sec:prelim}, we provide a comprehensive overview of related work in the field of homomorphic encryption and its application in outsourced databases. 
Section~\S\ref{sec:silca} presents the Silca and SilcaZ algorithms, including the advanced caching strategies, streaming techniques, and detailed security and complexity analysis.
Section~\S\ref{sec:system} details the practical implementation of Silca and SilcaZ with HElib as loadable functions in MySQL. 
In Section~\ref{sec:eval}, we present the experimental setup, datasets, and evaluation results, comparing the performance of Silca and SilcaZ with existing schemes,
such as CKKS~\cite{ckks}, BGV~\cite{bgv}, and Rache~\cite{otawose_sigmod23}. 
Finally, Section~\S\ref{sec:conc} concludes the paper, summarizing the contributions of this research and outlining potential avenues for future work.

\section{Preliminaries and Related Work}
\label{sec:prelim}

\subsection{Homomorphic Encryption}
The notion of \textit{homomorphism} originates from the study of algebraic groups~\cite{fraleigh_book03}, which are algebraic structures defined over nonempty sets. Formally, a group $G$ over a set $S$ is represented as a tuple $(G, \oplus)$, where $\oplus$ is a binary operator satisfying four axioms or properties, expressed as first-order logical formulas:
(i) For all $g, h \in S$, $g \oplus h \in S$.
(ii) There exists a unique element $u \in S$ such that for all $g \in S$, $(g \oplus u = g)$ and $(u \oplus g = g)$.
(iii) For every $g \in S$, there exists an element $h \in S$ such that $(g \oplus h = u)$ and $(h \oplus g = u)$. This element $h$ is often denoted as $-g$.
(iv) For all $g, h, j \in S$, $(g \oplus h) \oplus j = g \oplus (h \oplus j)$.
If we have another group $(H, \otimes)$ and a function $\varphi: G \rightarrow H$ such that for all $g_1, g_2 \in G$, $\varphi(g_1) \otimes \varphi(g_2) = \varphi(g_1 \oplus g_2)$, then we call the function $\varphi$ a homomorphism from $G$ to $H$. In other words, the function $\varphi$ preserves the group operation between the elements of $G$ when mapped to the corresponding elements in $H$.

\textit{Homomorphic encryption} (HE) is a specific type of encryption where certain operations between operands can be performed directly on the ciphertexts.
For example, if an HE scheme $he(\cdot)$ is additive,
then the plaintexts with $+$ operations can be translated into a homomorphic addition $\oplus$ on the ciphertexts.
Formally, if $a$ and $b$ are plaintexts, then the following holds:
\[
dec(he(a) \oplus he(b)) = a + b,
\]
where $dec$ denotes the decryption algorithm.

An HE scheme that supports addition is said to be \textit{additive}.
Popular additive HE schemes include Paillier~\cite{ppail_eurocrypt99},
which is an asymmetric scheme where a pair of public and private keys are used for encryption and decryption.
An HE scheme that supports multiplication is said to be \textit{multiplicative}.
Symmetria~\cite{symmetria_vldb20} is a recent scheme proposed in the database community,
which is multiplicative using a distinct scheme from the one for addition.
Other well-known multiplicative HE schemes include RSA~\cite{rsa} and ElGamal~\cite{elgamal_tit85}.
Similarly, a multiplicative HE scheme guarantees the following equality,
\[
dec (he(a) \otimes he(b)) = a \times b,
\]
where $\otimes$ denotes the homomorphic multiplication over the ciphertexts.

An HE scheme that supports both addition and multiplication is called a \textit{fully HE (FHE) scheme}.
This requirement should not be confused with specific addition and multiplication parameters, such as Symmetria~\cite{symmetria_vldb20} and NTRU~\cite{ntru}.
That is, the addition and multiplication must be supported homomorphically under the same scheme $he(\cdot)$:
\[\displaystyle
\begin{cases}
    dec( he(a) \oplus he(b) ) = a + b \\
    dec( he(a) \otimes he(b) ) = a \times b.
\end{cases}
\]
It turned out to be extremely hard to construct FHE schemes until Gentry~\cite{cgentry_stoc09} demonstrated such a scheme using lattice ideals.
Although lattice has been extensively studied in cryptography,
the combination of lattices and ring ideals is somewhat less explored;
nonetheless, Gentry showed that it is possible to construct an FHE scheme although the cost to maintain the multiplicative homomorphism is prohibitively high even with the so-called bootstrapping optimization,
which essentially applies decryption for every single multiplication operation between ciphertexts.

In the second generation of FHE schemes,
e.g., \cite{bgv,bfv,ckks},
the encryption efficiency has been greatly improved partially due to the removal of ideal lattices;
rather the new series of FHE schemes are based on the learning with error (LWE) or its variant ring learning with error (RLWE),
which have been proven to be as secure as hard lattice problems (e.g., through quantum or classical reduction).
The good news is that schemes that are built upon LWE or RLWE are significantly more efficient than the first-generation schemes.
However, there is still a wide gap between the state-of-the-art FHE cryptosystems and the desired performance.
Popular open-source libraries of FHE schemes include IBM HElib~\cite{helib}, 
Microsoft SEAL~\cite{sealcrypto}, and others.

\subsection{Performance Optimization for HE Schemes}

Hardware-based optimization of HE performance has been extensively studied in existing literature. For instance, researchers have explored optimization techniques \cite{nsam_micro21,dreis_vlsi20,ydoroz_tc15} for HE schemes. A recent article highlights the memory wall as the current performance bottleneck of HE \cite{decastro2021does}. However, our work is distinct from these studies as it focuses on purely algorithmic and software optimization.

In the context of databases, many works~\cite{btan_tdsc21,ilia_ppet21} focus on developing efficient comparison among ciphertexts because comparison is extensively invoked in database queries and joins.
The proposed algorithms of this work,
namely Silca and SilcaZ,
are orthogonal to those optimizations for tuple comparisons because Silca and SilcaZ aim to optimize the encryption performance rather than the algebraic opearation (i.e., comparison) among ciphertexts.

Another related research direction to our proposed caching-based optimization for HE schemes is called \textit{incremental cryptography},
which was first formally introduced in the 1990s \cite{mbellare_crypto94,mbellare_stoc95}, primarily from a theoretical standpoint. Recent work on incremental encryption schemes can be found in \cite{imironov_eurocrypt12,pananth_eurocrypt17,lkhati_sac18}. Incremental encryption has gained significant attention in current research, particularly for efficient data encoding in resource-constrained contexts like mobile computing \cite{fwang_fgcs21,gke_journal21,tbhatia_ccpe20}. However, to the best of our knowledge, no existing cryptosystem simultaneously supports both homomorphic encryption and incremental encoding while ensuring proven semantic security.

\subsection{Threat Model and Provable Security}

We assume the outsourced database servers are semi-honest.
When developing a new encryption scheme, it is crucial to assess its security level, preferably in a provable manner. One widely accepted approach, striking a balance between efficiency and security, is to consider the scenario where the adversary can launch a chosen-plaintext attack (CPA). In a CPA, the adversary can obtain the ciphertext of any arbitrarily chosen plaintext.
In practice, it is assumed that the adversary has limitations on the number of plaintext-ciphertext pairs they can obtain. Specifically, the adversary should only be able to access a polynomial number of such pairs, and their computational resources should be bounded by polynomial time. This assumption reflects the notion that adversaries should not have unlimited computational power or the ability to gather an excessive amount of information.
By making these assumptions, security analysis can be conducted under realistic constraints, allowing for a comprehensive evaluation of the encryption scheme's resilience against chosen-plaintext attacks. The goal is to design an encryption scheme that remains secure even when faced with an adversary who can selectively choose plaintexts and obtain their corresponding ciphertexts within the given limitations.

Ideally, even if the adversary $\mathcal{A}$ can obtain those extra pieces of information, $\mathcal{A}$ should not be able to make a \textit{distinguishably} better decision for the plaintexts than a random guess. This concept is quantified using the notion of a \textit{negligible function}.
A function $\mu(\cdot)$ is considered negligible if, for all polynomials $poly(n)$, the inequality $\mu(n) < \frac{1}{poly(n)}$ holds for sufficiently large values of $n$. 
In other words, the function $\mu(\cdot)$ decreases faster than the reciprocal of any polynomial as $n$ grows. Negligible functions are used to express the level of advantage an adversary can gain over a random guess, and they play a crucial role in the analysis of security proofs for cryptographic schemes.

\section{Singular Caching Algorithms for Homomorphic Encryption}
\label{sec:silca}

\subsection{High-Level Intuition}

The intuition behind the proposed singular caching for encrypted databases is very natural.
The main idea behind Silca is to optimize the caching process by reusing precomputed ciphertexts instead of generating them from scratch. This approach significantly reduces the overhead associated with encryption and decryption operations, making it practical for real-life applications.

Specifically, Silca introduces a concept called ``singular caching,'' where a single cached ciphertext is accessed only once during the online encryption stage. This is made possible by two essential factors. First, the underlying FHE scheme supports a type of multiplication operation that allows us to perform computations between plaintexts and ciphertexts efficiently. Second, the system utilizes an outsourced database server capable of offline and multithreading data updates, which is a common feature in modern production systems.

During the initialization phase, Silca constructs a circular buffer of precomputed ciphertexts using a technique called radix-based caching. This buffer is filled with encrypted random numbers generated based on a set of base building blocks. These ciphertexts are then used during the online encryption phase.

In the online encryption phase, Silca applies a randomization technique to select a specific ciphertext from the circular buffer. This selected ciphertext, called the ``mask,'' is combined with the plaintext message using element-wise multiplication. This process ensures that the resulting ciphertext does not reveal any information about the original plaintext.

By carefully manipulating the ciphertexts and leveraging the properties of the underlying FHE scheme, Silca achieves a constant-time overhead for caching operations, making it highly efficient in practical scenarios.

\subsection{Security Notions}

Before discussing the Silca encryption algorithm
we elaborate on the its security notion in terms of the security goal, threat model, and assumptions.

(i) Security Goal: The security goal of the Silca algorithm is to achieve IND-CPA (Indistinguishability under Chosen-Plaintext Attack) security. IND-CPA security ensures that an adversary cannot distinguish between the encryption of two different plaintext messages when provided with the corresponding ciphertexts, even if the adversary has the power to choose the plaintexts and obtain their encryptions.

(ii) Threat Model: The threat model for the Silca algorithm includes a passive adversary that has the ability to observe and interact with the system but does not have access to the secret key used for encryption. The adversary can perform chosen-plaintext attacks by providing plaintext messages and observing their corresponding ciphertexts. The adversary's goal is to determine which of two given plaintext messages was encrypted based on the observed ciphertexts.

(iii) Assumption: The security analysis of Silca assumes that the underlying base Fully Homomorphic Encryption (FHE) scheme used in Silca is CPA (Chosen-Plaintext Attack) secure. CPA security guarantees that the encryption produced by the base FHE scheme is indistinguishable from random under chosen-plaintext attacks. In other words, the encryption algorithm of the base FHE scheme hides any information about the plaintext message and provides semantic security.

By leveraging the assumed CPA security of the base FHE scheme, the Silca algorithm aims to provide IND-CPA security, which means that even though an adversary can observe the ciphertexts and choose the plaintexts, they cannot distinguish between the encryption of different plaintext messages.
It should be noted that the actual security of the Silca algorithm is dependent on the security properties of the underlying base FHE scheme and the correctness of its implementation.

\subsection{Silca Encryption Algorithm}

The Silca algorithm leverages a base fully homomorphic encryption (FHE) scheme to construct ciphertexts using the singular caching approach. During the initialization phase, a circular buffer array, $rp$, is created with each entry representing a circular buffer of ciphertexts. Random numbers $r_i$ are uniformly sampled and stored in the circular buffers for future use.

During the encryption phase, a randomization salt, $salt$, is generated. The element at index 0 of the circular buffer $rp[r_{salt}]$ is used as the $mask$. The element is then removed from the circular buffer and replaced with the encryption of $r_{salt}$ for streaming purposes. The ciphertext $ctxt$ is computed by element-wise multiplying the $mask$ and $ptxt$. Finally, $ctxt$ is multiplied with $\displaystyle \left( \frac{1}{r_{salt}} \right)$ to complete the encryption process.

\begin{algorithm}[!t]
\SetAlgorithmName{Algorithm}{}{}
\caption{Singular Caching (Silca)}\label{alg:silca}
\KwData{
A base FHE scheme applicable to floating-point numbers,
$FHE(KeyGen, Enc, Dec, \oplus, \otimes, \odot)$;
A plaintext message $ptxt$;
$\lfloor \log N \rfloor$ random numbers $r_i$ ($1 \le i \le \lfloor \log N \rfloor$) uniformly sampled from $[1, N]$,
where $N$ denotes the floor integer of the maximum of $ptxt$;
The data type of ciphertext $Ct$;
The length of the circular buffer $L$,
i.e., each $r_i$ has $L$ distinct entries;
}
\KwResult{
A ciphertext \textit{ctxt} s.t. $Dec_{sk}(ctxt) = ptxt$;
secret key $sk$ generated by $KeyGen$
}
 
\nonl \;
\nonl // Initialization (offline) \;
$array<circular\_buffer<Ct>, \lfloor \log N \rfloor> rp$ \;
\For{$j = 0; j < L; j++$}{
    \For{$idx = 1; idx \le \lfloor \log N \rfloor; idx++$}{
        $rp[idx].push\_back(Enc(r_{idx}))$ \;
    }
}
 
\nonl \;
\nonl // Encryption (online) \;
$salt \gets [1, \lfloor \log N \rfloor]$ // Randomization\;
$mask \coloneqq rp[r_{salt}][0]$\;
$rp[r_{salt}].pop\_front()$\;
$rp[r_{salt}].push\_back(Enc(r_{salt}))$ // Streaming\;
$ctxt \coloneqq mask \odot ptxt$\;
$\displaystyle ctxt \coloneqq ctxt \odot \left( \frac{1}{r_{salt}} \right)$\;
\end{algorithm}

Alg.~\ref{alg:silca} describes the algorithm.
The algorithm can be divided into two main phases: initialization (offline) and encryption (online).

During the initialization phase, the algorithm creates an array of circular buffers, denoted as $rp$, with $N+1$ elements. Each circular buffer in $rp$ corresponds to a specific random number $r_i$. The algorithm iterates over each circular buffer and fills it with encrypted values. The outer loop iterates $L$ times, and the inner loop iterates over the range $1$ to $\lfloor \log N \rfloor$. For each iteration, the algorithm encrypts the random number $r_{idx}$ using the base FHE encryption function, $Enc$, and stores the encrypted value in the corresponding circular buffer $rp[idx]$. Additionally, the circular buffer $rp[0]$ is populated with the encrypted value of the multiplicative identity, $Enc(1)$.

Once the initialization phase is complete, the algorithm proceeds to the encryption phase. It begins by creating a randomization array, $salt$, which contains the numbers $1$ to $\lfloor \log N \rfloor$. This array introduces randomization to the encryption process. The algorithm selects the first element of the circular buffer $rp[r_{salt}]$ and assigns it to the variable $mask$. This value serves as a randomization factor for encryption. The first element is then removed from $rp[r_{salt}]$ to maintain the circular buffer property. The algorithm updates $rp[r_{salt}]$ by pushing the encrypted value of $r_{salt}$ into the circular buffer, enabling streaming operations.

The ciphertext, denoted as $ctxt$, is computed by element-wise multiplication of the randomization factor $mask$ and the plaintext message $ptxt$. This operation is performed using the $\odot$ operator. Finally, $ctxt$ is further modified by element-wise multiplication with the reciprocal of $r_{salt}$ to ensure the correctness of the encryption.

\subsection{Correctness}

In this section, we aim to prove the correctness of the Silca algorithm, specifically regarding the decryption of the ciphertext generated by Silca. We will show that given a plaintext message $m$, the Silca decryption function $\text{Dec}_{\text{Silca}}(sk, c)$ correctly retrieves the original plaintext.

We define the following notations :
\begin{itemize}
\item $\text{FHE}$: The base Fully Homomorphic Encryption (FHE) scheme that Silca is built upon.
\item $\text{Enc}{\text{FHE}}(pk, m)$: The encryption function of the base FHE scheme, which takes a public key $pk$ and a plaintext message $m$ as inputs and produces a ciphertext.
\item $\text{Dec}{\text{FHE}}(sk, c)$: The decryption function of the base FHE scheme, which takes a secret key $sk$ and a ciphertext $c$ as inputs and produces the corresponding plaintext message.
\item $\text{Silca}$: The Silca algorithm that extends the functionality of the base FHE scheme with efficient ciphertext homomorphic operations.
\end{itemize}

To prove the correctness of Silca, we need to demonstrate that for any plaintext message $m$, the ciphertext generated by Silca, denoted as $c_{\text{Silca}}$, can be correctly decrypted back into the original plaintext using the Silca decryption function $\text{Dec}_{\text{Silca}}(sk, c_{\text{Silca}})$.

The Silca algorithm operates as follows:
\begin{enumerate}
    \item Random Selection: Silca randomly selects a ciphertext $c_{\text{rand}}$ from the FHE ciphertext space.
    \item Encryption: Silca computes the ciphertext $c_{\text{Silca}}$ as follows:

\[
c_{\text{Silca}} = c_{\text{rand}} \odot \frac{m}{\text{rand}}.
\]
\end{enumerate}

Now, to decrypt the Silca ciphertext $c_{\text{Silca}}$ and obtain the original plaintext message $m$, we apply the Silca decryption function $\text{Dec}_{\text{Silca}}(sk, c_{\text{Silca}})$ as follows:

\[
\text{Dec}_{\text{Silca}}(sk, c_{\text{Silca}}) = \text{Dec}_{\text{FHE}}\left(sk, c_{\text{rand}} \odot \frac{m}{\text{rand}} \right)
\]

By algebraic manipulation, we simplify the expression:

\[
\begin{aligned}
&\quad \text{Dec}_{\text{Silca}}\left(sk, c_{\text{Silca}}\right) \\
&= \text{Dec}_{\text{FHE}}\left(sk, c_{\text{rand}} \odot \frac{m}{\text{rand}} \right) \\
&= \text{Dec}_{\text{FHE}}\left(sk, \text{Enc}_{\text{FHE}}(pk, \text{rand}) \odot \frac{m}{\text{rand}} \right) \\
&= \text{Dec}_{\text{FHE}}\left(sk, \text{Enc}_{\text{FHE}}(pk, 1) \odot \text{rand} \odot \frac{m}{\text{rand}} \right) \\
&= \text{Dec}_{\text{FHE}}\left( sk, \text{Enc}_{\text{FHE}}(pk, 1) \odot m \right) \\
&= \text{Dec}_{\text{FHE}}\left( sk, \text{Enc}_{\text{FHE}}(pk, m) \right) \\
&= m.
\end{aligned}
\]

Therefore, we have shown that the Silca decryption function correctly retrieves the original plaintext message $m$ from the Silca ciphertext $c_{\text{Silca}}$, ensuring the correctness of the Silca algorithm.

\subsection{Parameterization}

Before proving the security of Silca,
we conduct some parameter analysis of the length of the circular buffer,
$L$.
We want to gain a better understanding of $L$ in practice.
Let $\phi$ denote the performance ratio of $Enc_{\text{FHE}}$ and $Enc_{\text{Silca}}$;
that is, their execution time satisfies the following
\[
T(\text{Enc}_{\text{FHE}}) = \phi \cdot T(\text{Enc}_{\text{Silca}}).
\]
One property of picking $L$ is to ensure that the cached ciphertext should be sufficient for the given set of plaintexts,
whose cardinality is denoted by $n$.

To ensure the cached ciphertexts are sufficient,
the following inequality holds
\[\displaystyle
n \cdot T(\text{Enc}_{\text{Silca}}) \ge \left (n - L\cdot \lfloor \log N \rfloor \right) \cdot T(\text{Enc}_{\text{FHE}}),
\]
which gives
\[
L \ge \frac{(\phi - 1)n}{\phi \lfloor \log N \rfloor}.
\]
While theoretically $L$ should be set as $\mathcal{O}(n)$,
in practice the cached ciphertexts can be efficiently calculated by multiple threads (to be reported in~\S\ref{sec:eval}, ref. Fig.~\ref{fig:overhead}).

The above inquality also implies that the total number of ciphertexts is bounded by
\[
n \le\frac{\phi \lfloor \log N \rfloor}{\phi - 1} \cdot L.
\]
One potential method to extend this limit is to leverage the additive caching (e.g., Rache~\cite{otawose_sigmod23}) to feed the circular buffer by adding two or more existing ciphertexts instead of constructing a new ciphertext factor from scratch (Line 10 of Alg.~\ref{alg:silca}).
Because every ciphertext factor can be expressed by an arbitrary pair of existing ciphertexts in a linear function,
i.e., $ctxt_2 = a\odot ctxt_1 + b\odot ctxt_2$, 
where $(a,b) \in \mathbb{R}^2$,
there are overall $\displaystyle \binom{L\lfloor \log N \rfloor}{2}$ possible replacements for a specific ciphertext factor in the circular buffer.
This implies that $n$ is bounded by a much larger number
\[\displaystyle
n \le\frac{\phi \lfloor \log N \rfloor}{\phi - 1} \cdot \binom{L\lfloor \log N \rfloor}{2} \cdot L.
\]
In practice, if we assume $\phi$ and $L$ are sufficiently large,
then a simpler upper bound of $n$ would be
$\displaystyle \lfloor \log N \rfloor ^ 3 \cdot L^3$.

\subsection{Semantic Security}

\begin{proposition}
Assuming that the base FHE scheme used in Silca is IND-CPA secure;
Silca provides IND-CPA security.    
\end{proposition}

\begin{proof}
To demonstrate IND-CPA security, we need to show that for any pair of plaintext messages $m_0$ and $m_1$ of the same length, an adversary cannot distinguish between the encryption of $m_0$ and $m_1$ when provided with the ciphertext.

Assume we have an adversary $\mathcal{A}$ that tries to distinguish between the encryption of $m_0$ and $m_1$. We construct a CPA game where $\mathcal{A}$ interacts with a challenger and attempts to guess the bit $b$ (0 or 1) corresponding to the encryption of either $m_0$ or $m_1$.

The CPA game proceeds as follows.
The challenger generates a random bit $b$.
The challenger selects the corresponding plaintext message $m_b$ and applies the Silca encryption algorithm to obtain the ciphertext $ctxt$.
The challenger provides $ctxt$ to $\mathcal{A}$.
$\mathcal{A}$ performs any number of operations (queries, computations, etc.) using the ciphertext $ctxt$ and outputs its guess for the bit $b$.
The challenger compares $\mathcal{A}$'s guess with the actual value of $b$ and outputs 1 if they match, and 0 otherwise.
To prove IND-CPA security, we need to show that for any efficient adversary $\mathcal{A}$, the advantage $\mathcal{A}$ has in distinguishing $m_0$ and $m_1$ is negligible. In other words, we aim to prove that the probability of $\mathcal{A}$ guessing the bit $b$ correctly is close to 0.

Let us assume we have an adversary $\mathcal{A}$ that can break Silca with non-negligible advantage, meaning that $\mathcal{A}$ can distinguish between encryptions of different plaintext messages. 
That is,
\[
\text{Adv}_{\mathcal{A}} > \frac{1}{2} + \frac{1}{poly(\lambda)},
\]
where $\lambda$ denotes the security parameter.
We aim to construct an adversary $\mathcal{B}$ that uses $\mathcal{A}$ to break the base FHE scheme.

The adversary $\mathcal{B}$ operates as follows:
\begin{enumerate}
    \item $\mathcal{B}$ receives the challenge from the base FHE scheme, which consists of two plaintext messages $m_0$ and $m_1$.
    \item $\mathcal{B}$ generates the encryptions of $m_0$ and $m_1$ using the base FHE scheme to obtain ciphertexts $C_0$ and $C_1$, respectively.
    \item $\mathcal{B}$ queries $\mathcal{A}$ with $C_0$ and $C_1$ to obtain the guessed bit $\hat{b}$.
    \item $\mathcal{B}$ outputs $\hat{b}$ as the final guess for the bit in the base FHE scheme challenge.
\end{enumerate}

To analyze the advantage of $\mathcal{B}$ in breaking the base FHE scheme, we consider the following cases.

Case 1: 
If the $c_{\text{rand}}$ happens to be $c_1$,
which means $\mathcal{A}$ is ``super lucky'' to have select the multiplicative identity 1 as the randomness.
Note that $\mathcal{A}$ cannot control the randomness procedure as per the definition of the experiment.
The chance for $\mathcal{A}$ to be in this case is 
$\frac{1}{2^\lambda}$.
When $\mathcal{B}$ is simulated on such an $\mathcal{A}$,
$\mathcal{B}$ has a slightly smaller chance to win the game than $\mathcal{A}$ by a negligible function.

Case 2: 
Otherwise, $\mathcal{B}$ and $\mathcal{A}$ perceive the information.
In this case, the chance for $\mathcal{B}$ to win the game is exactly the same as $\mathcal{A}$.

Combining both above cases,
the overall advantage for $\mathcal{B}$ to win the experiment is
\[
\text{Adv}_{\mathcal{B}} > \frac{1}{2} + \frac{1}{poly(\lambda)} - \frac{1}{2^\lambda},
\]
which implies that FHE can be broken by $\mathcal{B}$ with a non-negligible advantage than a random guess.

The final piece we need to verify is that whether the reduction can be done \textit{efficiently}.
That is, whether we can complete the transition of views for $\mathcal{A}$ and $\mathcal{B}$ in polynomial time.
This is clearly the case by checking Lines 7--11 of Alg.~\ref{alg:silca}.
In fact, the simulation completes in constant time $\mathcal{O}(1)$.
\end{proof}

\subsection{Time Complexity}

The time complexity of the Silca algorithm can be analyzed as follows:

\paragraph{Initialization Phase}
Step 2: The outer loop runs $L$ times, and the inner loop runs $\lfloor \log N \rfloor$ times. Hence, the time complexity of the initialization phase is $O(L \cdot \log N)$.

\paragraph{Encryption Phase}
Step 1: Generating the randomization salt requires constant time, so its time complexity is $O(1)$.
Steps 2-4: Accessing and modifying the circular buffer take constant time, resulting in a time complexity of $O(1)$.
Steps 5-6: Element-wise multiplication and division operations on the ciphertexts can be performed in constant time due to the homomorphic property of the underlying FHE scheme. Therefore, the time complexity of these steps is also $O(1)$.

Overall, the time complexity of the Silca algorithm is dominated by the offline stage, which is $O(L \cdot \log N)$. The real-time encryption phase has a constant time complexity of $O(1)$.


\subsection{Space Complexity}

The space complexity of the Silca algorithm can be analyzed as follows.
\begin{itemize}
\item Initialization Phase.
The circular buffer array, $rp$, has a size of $N+1$. Each entry in the array is a circular buffer of ciphertexts. Therefore, the space complexity of the initialization phase is $O(N \cdot \text{sizeof}(Ct) \cdot L)$, where $\text{sizeof}(Ct)$ represents the size of the ciphertext data type.

\item Encryption Phase.
The space complexity of the encryption phase is mainly determined by the storage of the randomization salt, $salt$, which requires constant space as it is a single value.
The space complexity of the other variables, such as $mask$, $ctxt$, and temporary variables used in the FHE operations, can be considered constant.
\end{itemize}

Overall, the dominant factor in the space complexity of the Silca algorithm is the circular buffer array, $rp$, which is responsible for storing the precomputed ciphertexts during the initialization phase. The space complexity of the Silca algorithm can be expressed as $O(N \cdot \text{sizeof}(Ct) \cdot L)$.


\subsection{SilcaZ: Silca for Integers}

SilcaZ is an integer version of the Silca algorithm that operates on integer values using a base fully homomorphic encryption (FHE) scheme. The algorithm aims to encrypt a plaintext message $ptxt$ and generate a ciphertext $ctxt$ such that the decryption of $ctxt$ recovers the original $ptxt$.
The properties are similar to Silca so we will not discuss them in detail.

We present SilcaZ in Alg.~\ref{alg:silcaz}.
The algorithm begins with an initialization phase where circular buffers, denoted by $rp$, are created. There are $\lfloor \log N \rfloor$ circular buffers, each storing ciphertexts of the random numbers $r_i$. These random numbers are uniformly sampled from the interval $[2, N-1]$, where $N$ represents the plaintext modulus and is chosen to be larger than the maximum value of $ptxt$. This ensures the ciphertexts remain within the range of representable integers.

\begin{algorithm}[!t]
\SetAlgorithmName{Algorithm}{}{}
\caption{SilcaZ (Integer Version of Silca)}\label{alg:silcaz}
\KwData{
A base FHE scheme applicable to integers,
$FHE(KeyGen, Enc, Dec, \oplus, \otimes, \odot)$;
A plaintext message $ptxt$ represented as an integer;
$\lfloor \log N \rfloor$ random numbers $r_i$ ($1 \le i \le \lfloor \log N \rfloor$) uniformly sampled from $[2, N-1]$,
where $N$ represents the plaintext modulus that is larger than the maximum value of $ptxt$;
The data type of ciphertext $Ct$;
The length of the circular buffer $L$,
i.e., each $r_i$ has $L$ distinct entries;
}
\KwResult{
A ciphertext \textit{ctxt} such that $Dec_{sk}(ctxt) = ptxt$;
secret key $sk$ generated by $KeyGen$
}

\nonl \;
\nonl // Initialization (offline) \;
$array<circular\_buffer<Ct>,\lfloor \log N \rfloor> rp$ \;
\For{$j = 0; j < L; j++$}{
    \For{$idx = 1; idx \le \lfloor \log N \rfloor; idx++$}{
        $rp[idx].push\_back(Enc(r_{idx}))$ \;
    }
}

\nonl \;
\nonl // Encryption (online) \;
$salt \gets [1, \lfloor \log N \rfloor]$ // Randomization\;
$mask \coloneqq rp[r_{salt}][0]$\;
$rp[r_{salt}].pop\_front()$\;
$rp[r_{salt}].push\_back(Enc(r_{salt}))$ // Streaming\;
$ctxt \coloneqq mask \odot ptxt$\;
$\displaystyle ctxt \coloneqq ctxt \odot \left( r_{salt}^{-1} \mod N \right)$\;
\end{algorithm}

During the encryption phase, a randomization factor $salt$ is selected from the interval $[1, \lfloor \log N \rfloor]$. The circular buffer $rp[r_{salt}]$ is accessed to obtain the first ciphertext element, which is assigned to the variable $mask$. The first element is then removed from $rp[r_{salt}]$ and replaced with a new ciphertext obtained by encrypting $r_{salt}$ using the FHE scheme.

Next, the plaintext message $ptxt$ is multiplied element-wise with the $mask$ ciphertext, resulting in the ciphertext $ctxt$. Finally, the multiplicative inverse of $r_{salt}$ modulo $N$ is computed, and $ctxt$ is multiplied by this inverse to further secure the encryption.
The algorithm outputs the ciphertext $ctxt$, which can be decrypted using the corresponding secret key $sk$ generated by the FHE scheme, thereby recovering the original plaintext message $ptxt$.

It should be noted that because $N$ is selected as a (large) prime number,
the inverse of $r_{salt}$ always exists in group $\mathbb
{Z}_{N}$.
Therefore, the notion $r^{-1}_{salt}$ is well defined in Line 12 of Alg.~\ref{alg:silcaz}.

All of the SilcaZ properties, 
such as security guarantee and complexity analysis,
follow those of Silca.
We, therefore, skip the discussion of SilcaZ properties in this paper.

\section{Silca/Z Loadable Functions in MySQL}
\label{sec:system}

We have implemented both Silca and SilcaZ with C++ on top of the CKKS and BGV schemes in HElib~\cite{helib}. 
Both the Silca and SilcaZ schemes are compiled into MySQL 8.0 as loadable functions.
As of the writing of this paper, there are overall 2,655 lines of code.

\subsection{Implementing Silca with CKKS}

\subsubsection{Software Environment}

The system implementation of SilcaZ involves parameterization using various libraries and components. The key parameters and libraries used in the implementation are described below:

\begin{itemize}
\item \textbf{MySQL and GMP Libraries}: The system is designed to work with the MySQL server, and it includes the necessary header files (\texttt{<mysql.h>} and \texttt{<m_string.h>} or \texttt{<string.h>}). Additionally, the GNU Multiple Precision Arithmetic Library (GMP) is utilized for high-precision arithmetic operations.
\item \textbf{Boost Library}: The Boost library is used to leverage functionalities such as circular buffers, string formatting, and date/time operations. It includes the \texttt{<boost/circular_buffer.hpp>} and \texttt{<boost/format.hpp>} headers, as well as the \texttt{boost::posix_time} namespace for date/time operations.
\item \textbf{HElib Library}: The system integrates the HElib library, which provides support for homomorphic encryption schemes. The \texttt{<helib/helib.h>} header is included, and the necessary namespaces (\texttt{helib} and \texttt{std}) are utilized.
\item \textbf{OpenMP Library}: OpenMP is used for multi-threading support in the system. The \texttt{<omp.h>} header is included to enable parallel execution of certain operations.
\end{itemize}

\subsubsection{Offline Caching}

The Silca algorithm incorporates an offline caching mechanism to optimize computation during the online phase. Offline caching involves precomputing and storing a set of ciphertexts, reducing redundant computations and enhancing overall performance. The process can be outlined as follows:

\textbf{Thread Configuration}: The offline caching process takes advantage of parallel execution. By configuring the number of threads (\texttt{n_thread}), Silca maximizes computational efficiency.

\textbf{Initialization}: Relevant parameters and variables are initialized, including data structures and memory allocation for the offline buffer.

\textbf{Offline Computation}: The \texttt{ckks_offline()} function executes the offline caching process. It involves iterating over a range of tasks, where each task corresponds to the computation of a specific ciphertext used during the online phase.

\textbf{Thread Assignment}: Tasks are assigned to different threads to distribute the workload evenly. Each thread focuses on its assigned tasks, skipping those belonging to other threads. This efficient thread assignment ensures optimal utilization of computational resources.

\textbf{Ciphertext Generation}: For each assigned task, a plaintext array is generated to represent the data to be encrypted. The generated plaintext is then encrypted using the public key, resulting in a corresponding ciphertext.

\textbf{Ciphertext Storage}: The computed ciphertext is stored in the offline buffer associated with the thread. This buffer acts as a cache, storing precomputed ciphertexts for efficient retrieval during the online phase.

\textbf{Optional Logging}: An optional logging mechanism can be enabled to track the progress of the offline computation. This provides insights into the status and performance of the offline caching process.

\textbf{Time Measurement}: The duration of the offline caching process is measured to evaluate its efficiency. The elapsed time is recorded and logged as the offline caching time.

The offline caching step in Silca significantly reduces redundant computations during the online phase, leading to improved efficiency and reduced computation time. By precomputing and storing ciphertexts, Silca optimizes the utilization of computational resources and enhances the overall performance of the algorithm.

\subsubsection{Silca on MySQL}

The \texttt{silca_ckks} function is an optimized encryption method based on the CKKS scheme. It offers faster encryption compared to traditional methods.
The function takes a plaintext value of floating-point number as input and returns the corresponding encrypted ciphertext.

The \texttt{silca_ckks()} function performs the following steps:

\textbf{Initialization:} The necessary variables and context for encryption are set up.

\textbf{Timer Start:} A timer is initiated to measure the encryption time.

\textbf{Random Factor:} A random factor is generated to introduce randomness into the encryption process. This factor is obtained from a pre-defined set of ciphertexts.

\textbf{Randomization:} A random salt value is generated using a random number generator. This salt value is used to divide the plaintext value, introducing additional randomness.

\textbf{Fast Encryption:} The plaintext value is multiplied by the random factor and the salt value to perform fast encryption. This technique significantly accelerates the encryption process.

The \texttt{silca_ckks()} function incorporates optimizations to enhance the efficiency of the CKKS encryption process. By reducing computational overhead and leveraging fast encryption techniques, it provides improved performance in the context of MySQL.

\subsection{Implementing SilcaZ with BGV}

\subsubsection{SilcaZ Parametrization}

The parametrization of SilcaZ involves setting various configuration options and initializing the necessary objects. 
We start by defining the parameters for the base BGV scheme. These parameters include:

\begin{itemize}
    \item The plaintext prime modulus $p$, which determines the range of valid plaintext values.
    \item The cyclotomic polynomial $m$, which defines the structure of the polynomial ring.
    \item The Hensel lifting parameter $r$, which controls the precision of the computations.
    \item The number of bits $bits$ in the modulus chain, which affects the security level and computation efficiency.
    \item The number of columns $c$ in the Key-Switching matrix, which impacts the efficiency of key-switching operations.
\end{itemize}

Using these parameters, we construct the global BGV context by invoking the \texttt{ContextBuilder} with the specified values. This creates an instance of \texttt{helib::Context}.

Next, we generate a secret key associated with the BGV context using \texttt{helib::SecKey}. The secret key is then generated using \texttt{GenSecKey()}.
We also initialize the public key (\texttt{public\_key}) by setting it equal to the secret key.
To facilitate encryption and decryption operations, we create an empty ciphertext object (\texttt{ctxt}) and an initialized plaintext object (\texttt{ptxt\_ary}) using the BGV context.

Moving on, we consider the circular caching array of HE ciphertexts. This array consists of various configurations and constants:

\begin{itemize}
    \item \texttt{RADIX\_BASE} represents the base of the radix for digit encoding.
    \item \texttt{RADIX\_MAX\_EXP} specifies the maximum exponent of the radix.
    \item \texttt{SZ\_CBUFF} determines the size of the circular buffer.
    \item \texttt{SZ\_RADIX} denotes the number of digits in the radix.
    \item \texttt{SZ\_DIGIT\_INT} and \texttt{SZ\_DIGIT\_FRA} represent the sizes of the integer and fraction parts, respectively.
    \item \texttt{SZ\_DIGIT\_RACHE} indicates the number of digits for the Rache part.
    \item \texttt{DIGIT\_MIN} and \texttt{DIGIT\_MAX} define the minimum and maximum plaintext values.
    \item \texttt{EPSILON} represents the precision epsilon between two real numbers.
    \item \texttt{DIGIT\_MOD} is the modular value for integer division.
    \item \texttt{THD\_MAX} specifies the maximum number of threads.
    \item \texttt{CACHED\_CTXT\_MAX} defines the maximum number of cached ciphertexts.
\end{itemize}

Additionally, we keep track of various statistics using variables such as \texttt{tot\_bgv\_vanilla}, \texttt{tot\_bgv\_rache}, \texttt{tot\_bgv\_streche}, and \texttt{tot\_bgv\_silca}.
The circular buffers \texttt{cb0}, \texttt{cba}, \texttt{cba\_int}, and \texttt{offline} are utilized for caching ciphertexts and integer values, ensuring efficient computation.

Furthermore, we define the structure \texttt{cb3} to hold a circular buffer of ciphertexts and \texttt{cb\_int} to store a circular buffer of integers. These structures are used in arrays \texttt{cba} and \texttt{cba\_int}, respectively.

\subsubsection{SilcaZ on MySQL}

The SilcaZ BGV implementation involves the following steps.
A random exponent (\texttt{exp}) is generated within the range of the Rache digits (\texttt{SZ_DIGIT_RACHE}). This exponent is used to access the corresponding ciphertext (\texttt{c}) from the \texttt{rache} vector.

To perform the encryption, we compute the inverse of the exponent modulo \texttt{p}, where \texttt{p} represents the plaintext prime modulus. This inverse (\texttt{inv}) is multiplied by the plaintext (\texttt{ptxt}) and the ciphertext (\texttt{c}), resulting in the encrypted value.
We accumulate the encryption time in the \texttt{tot_bgv_silca} variable, which keeps track of the total SilcaZ BGV encryption time.

At this point, the encryption process is complete. Depending on the requirements, additional actions such as logging or returning the encrypted ciphertext can be implemented.

\subsection{Deployment on MySQL Loadable Functions}

\subsubsection{MySQL Server}

In order to deploy Silca loadable functions in MySQL, we follow the steps outlined below:

First, we need to set the correct paths for MySQL includes and plugins. We define \texttt{PATH_MYSQL_INC} as the path for MySQL includes and \texttt{PATH_MYSQL_PLUGIN} as the directory for plugins.

Next, we compile the \texttt{silca_bgv.cpp} source file for Silca-BGV. During compilation, we include the necessary flags and libraries. This generates a shared object file named \texttt{silca_bgv.so}. We then move this file to the MySQL plugins directory.

Similarly, for Silca-CKKS, we compile the \texttt{silca_ckks.cpp} source file with the required flags and libraries. This produces a shared object file called \texttt{silca_ckks.so}, which is also moved to the MySQL plugins directory.

To ensure the availability of the updated functions, it is essential to restart MySQL. This step refreshes the MySQL environment and allows the newly added Silca plugins to be loaded.

By following these steps, we successfully deploy Silca-BGV and Silca-CKKS as loadable functions in MySQL. This deployment empowers us to leverage Silca/Z's advanced capabilities for secure computation tasks within the MySQL ecosystem.

\subsubsection{Python Client}

The Python client code presented below demonstrates the interaction with the MySQL database using the PyMySQL library. The client performs various tasks, such as initializing tables, creating and dropping functions, executing queries, and verifying results.

To establish a connection with the MySQL server, the client code specifies the host, user, password, and other connection parameters. The pymysql.connect() function is used to create the connection object.

The client code utilizes the cursor object to execute SQL statements and retrieve query results. The cursor class is set to pymysql.cursors.DictCursor, which returns query results as dictionaries.

The client code defines several helper functions:

\begin{itemize}
    \item bytes_to_int() and int_to_bytes(): These functions convert between bytes and integers.
    \item init_table(): This function creates a table named \textit{ctxt} if it doesn't already exist.
    \item init_function() and drop_function(): These functions create and drop the \textit{silca} and \textit{silcaZ} functions, respectively.
    \item drop_table(): This function drops the \textit{ctxt} table if it exists.
\end{itemize}

The main part of the client code includes two test functions:

\begin{itemize}
    \item test_silca(): This function shows an example of using the silca_ckks() function. For example, it executes a SELECT query on the `view_lineitem' table, applying the silca() function to the `l_extendedprice' attribute.

    \item test_silcaz(): Another function is similarly defined for testing the SilcaZ-BGV scheme.
\end{itemize}

After executing the necessary functions, the client code outputs the contents of a log file /tmp/log_silca.txt. Additionally, there is a commented-out section that can be used to check the ciphertext stored in a binary file /tmp/ctxt.json.

By running this Python client code, one can interact with the MySQL database, execute custom functions, and perform silca/Z operations on the data stored in the tables.

\section{Evaluation}
\label{sec:eval}

\subsection{Hardware Specification}

The hardware specification of the system used in this study is as follows. The system is hosted at the TACC site and belongs to the ChameleonCloud~\cite{katek_atc20}. It is built on the x86\_64 platform and features 2 CPUs with a total of 256 threads. The system is equipped with 256 GiB of RAM, providing ample memory capacity for computational tasks. The node type is compute\_zen3.
The chassis of the system is manufactured by Dell Inc. and has the model name PowerEdge R6525.

The processor of the system is an AMD EPYC 7763 64-Core Processor. It operates at a clock speed of 2.45 GHz and supports the following cache configurations: L1d cache, L1i cache, L2 cache with a capacity of 32,768,000 bytes, and L3 cache with a capacity of 262,144,000 bytes.

\subsection{Software and Dependent Libraries}


We are allocated the c02-14 instance by the Chameleon Cloud~\cite{katek_atc20}, operating on Ubuntu 20.04 LTS. The software and library versions used in our system are as follows: MySQL database version 8.0.33, HElib version 2.2.2, g++ version 9.4.0, and Boost C++ library version 1.71.

In addition, important dependencies include Number theory library (NTL) version 11.4.3 and Multiple-precision arithmetic library (GMP) version 6.2.0.

The source code files are compiled with the following flags and options: \textit{-fPIC} (generating position-independent code required for loadable functions), \textit{-fopenmp} (enabling OpenMP support for thread-level parallelization), and \textit{-std=C++17} (setting the C++ language standard to C++17).

These specific configurations and versions ensure compatibility and enable the desired features and functionality in our system implementation.

\subsection{Datasets}

Table \ref{tbl:benchmark} presents a list of datasets to be evaluated with. The table consists of seven columns, describing the dataset name, the number of tuples, the data type, the minimal value, the maximal value, the average value, and the standard variance.

\begin{table*}[!t]
  \centering
  \caption{List of datasets}
  \label{tbl:benchmark}
  \renewcommand{\arraystretch}{1.2}
  \rowcolors{2}{gray!20}{white}
  \begin{tabular}{l c c c c c c}
    \toprule
    \rowcolor{gray!50}
    \textbf{Dataset Name} & \textbf{Number of Tuples} & \textbf{Data Type} & \textbf{Minimal Value} & \textbf{Maximal Value} & \textbf{Average Value} & \textbf{Standard Variance} \\
    \midrule
    \rowcolor{gray!10}
    Covid19 & 341 & Integer & 123,021 & 2,309,884 & 1,063,465.029 & 570,009.089 \\
    Bitcoin & 1,086 & Float & 274,252.698 & 9,999,999.999 & 7,412,197.895 & 3,472,109.034 \\
    \rowcolor{gray!10}
    hg38 & 34,424 & Integer & 1 & 360 & 10.915 & 9.891 \\
    P\_Size & 200,000 & Integer & 1 & 50 & 25.427 & 14.441 \\
    \rowcolor{gray!10}
    P\_RetailPrice & 200,000 & Float & 901.00 & 2,098.99 & 1,499.49 & 294.673 \\
    O\_TotalPrice & 1,500,000 & Float & 857.71 & 555,285.16 & 151,219.537 & 88,621.401 \\
    \rowcolor{gray!10}
    L\_ExtendedPrice & 6,001,215 & Float & 901.00 & 104,949.50 & 38,255.138 & 23,300.436 \\
    \bottomrule
  \end{tabular}
\end{table*}

The first dataset, Covid19~\cite{covid19data}, is represented by 341 tuples and consists of integer data type. The minimal value in this dataset is 123,021, while the maximal value reaches 2,309,884. The average value of the Covid19 dataset is 1,063,465.029, with a standard variance of 570,009.089.

The second dataset, Bitcoin~\cite{bitcoin_trade}, contains 1,086 tuples and utilizes the float data type. The minimal value in this dataset is 274,252.698, while the maximal value is 9,999,999.999. The average value of the Bitcoin dataset is 7,412,197.895, accompanied by a standard variance of 3,472,109.034.

The third dataset, hg38~\cite{hg_data}, comprises 34,424 tuples of integer data type. The minimal value in this dataset is 1, and the maximal value is 360. The average value for the hg38 dataset is 10.915, with a standard variance of 9.891.

The remaining four datasets are from the TPC-H benchmark~\cite{tpch3}.

The fourth dataset, P\_Size, consists of 200,000 tuples with integer data type. The minimal value in this dataset is 1, while the maximal value is 50. The average value for the P\_Size dataset is 25.427, and the standard variance is 14.441.

The fifth dataset, P\_RetailPrice, contains 200,000 tuples utilizing the float data type. The minimal value in this dataset is 901.00, while the maximal value is 2,098.99. The average value of the P\_RetailPrice dataset is 1,499.49, accompanied by a standard variance of 294.673.

The sixth dataset, O\_TotalPrice, comprises 1,500,000 tuples of float data type. The minimal value in this dataset is 857.71, and the maximal value is 555,285.16. The average value for the O\_TotalPrice dataset is 151,219.537, with a standard variance of 88,621.401.

The final dataset, L\_ExtendedPrice, contains 6,001,215 tuples utilizing the float data type. The minimal value in this dataset is 901.00, while the maximal value is 104,949.50. The average value of the L\_ExtendedPrice dataset is 38,255.138, and the standard variance is 23,300.436.



\subsection{Performance of Cryptgraphic Primitives}

\subsubsection{CKKS}

Table \ref{tbl:func_ckks} presents a detailed comparison of the computational cost of key CKKS algorithms (functionalities) implemented as MySQL loadable functions. The table consists of three columns, describing the CKKS functionality, the raw time taken for computation, and the relative speed compared to the \textit{CKKS\_Enc} functionality.

\begin{table}[!t]
  \begin{center}
    \caption{Comparing the computational cost of key CKKS~\cite{ckks} algorithms (i.e., functionalities) in MySQL loadable functions.}
    \label{tbl:func_ckks}
    \begin{tabular}{l r r}
\toprule
    CKKS Functionality & Raw Time & Relative Speed \\
\midrule
    \textit{CKKS\_Enc} & 2.47 ms & 1$\times$ \\
    \textit{CKKS\_EvalAdd} & 0.13 ms & 19$\times$ \\
    \textit{CKKS\_EvalAddPlain} & 0.11 ms & 22$\times$ \\
    \textit{CKKS\_EvalMulPlain} & 0.07 $\mu$s & 35k$\times$ \\
\bottomrule
    \end{tabular}
  \end{center}
\end{table}

The first functionality, \textit{CKKS\_Enc}, demonstrates a raw time of 2.47 milliseconds. It serves as the baseline for comparison, denoted as 1$\times$. This functionality represents the encryption process in the CKKS scheme.

The second functionality, \textit{CKKS\_EvalAdd}, exhibits a significantly reduced computation time of 0.13 milliseconds. It achieves a relative speed of 19$\times$ compared to the \textit{CKKS\_Enc} functionality. The \textit{CKKS\_EvalAdd} functionality is responsible for performing addition operations in the CKKS scheme.

The third functionality, \textit{CKKS\_EvalAddPlain}, shows an even further reduction in computation time, taking only 0.11 milliseconds. It achieves a relative speed of 22$\times$ compared to \textit{CKKS\_Enc}. The \textit{CKKS\_EvalAddPlain} functionality combines the addition of encrypted values with plaintext values in the CKKS scheme.

The fourth functionality, \textit{CKKS\_EvalMulPlain}, demonstrates the most efficient computation time of 0.07 microseconds, showcasing the highest relative speed in the table, reaching 35,000$\times$ compared to \textit{CKKS\_Enc}. The \textit{CKKS\_EvalMulPlain} functionality performs multiplication operations involving encrypted values and plaintext values in the CKKS scheme.


\subsubsection{BGV}

Table \ref{tbl:func_bgv} provides a detailed comparison of the computational cost of key BGV algorithms (functionalities) implemented as MySQL loadable functions. The table consists of three columns, describing the BGV functionality, the raw time taken for computation, and the relative speed compared to the \textit{BGV\_Enc} functionality.

\begin{table}[!t]
  \begin{center}
    \caption{Comparing the computational cost of key BGV~\cite{bgv} algorithms (i.e., functionalities) in MySQL loadable functions.}
    \label{tbl:func_bgv}
    \begin{tabular}{l r r}
\toprule
    BGV Functionality & Raw Time & Relative Speed \\
\midrule
    \textit{BGV\_Enc} & 48.20 ms & 1$\times$ \\
    \textit{BGV\_EvalAdd} & 0.19 ms & 254$\times$ \\
    \textit{BGV\_EvalAddPlain} & 1.94 ms & 25$\times$ \\
    \textit{BGV\_EvalMulPlain} & 0.11 $\mu$s & 438k$\times$ \\
\bottomrule
    \end{tabular}
  \end{center}
\end{table}

The first functionality, \textit{BGV\_Enc}, exhibits a raw time of 48.20 milliseconds, serving as the baseline for comparison, denoted as 1$\times$. This functionality represents the encryption process in the BGV scheme.

The second functionality, \textit{BGV\_EvalAdd}, demonstrates a significantly reduced computation time of 0.19 milliseconds. It achieves a remarkable relative speed of 254$\times$ compared to the \textit{BGV\_Enc} functionality. The \textit{BGV\_EvalAdd} functionality is responsible for performing addition operations in the BGV scheme.

The third functionality, \textit{BGV\_EvalAddPlain}, shows a slightly increased computation time of 1.94 milliseconds. It achieves a relative speed of 25$\times$ compared to \textit{BGV\_Enc}. The \textit{BGV\_EvalAddPlain} functionality combines the addition of encrypted values with plaintext values in the BGV scheme.

The fourth functionality, \textit{BGV\_EvalMulPlain}, exhibits the most efficient computation time of 0.11 microseconds, showcasing the highest relative speed in the table, reaching 438,000$\times$ compared to \textit{BGV\_Enc}. The \textit{BGV\_EvalMulPlain} functionality performs multiplication operations involving encrypted values and plaintext values in the BGV scheme.



\subsection{Caching Overhead}

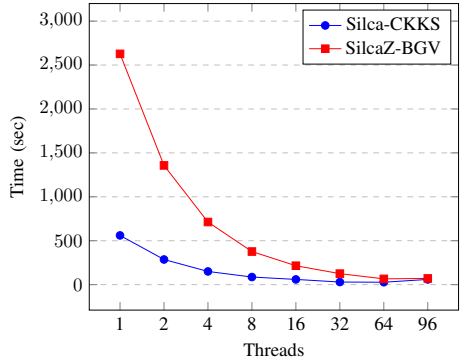
\begin{figure}
  \centering
    \resizebox{0.7\linewidth}{!}{ 

\begin{tikzpicture}
\begin{axis}[
    xlabel={Threads},
    ylabel={Time (sec)},
    legend style={at={(0.98,0.98)}, anchor=north east},
    symbolic x coords={1, 2, 4, 8, 16, 32, 64, 96},
    xtick=data,
    ytick={0, 500, 1000, 1500, 2000, 2500, 3000},
    ymax=3200,
    ymajorgrids=true,
    grid style=dashed,
]

\addplot[color=blue, mark=*] coordinates {
    (1, 560.3)
    (2, 286.2)
    (4, 150.1)
    (8, 86.5)
    (16, 59.4)
    (32, 29.5)
    (64, 27.3)
    (96, 59.3)
};

\addplot[color=red, mark=square*] coordinates {
    (1, 2627.3)
    (2, 1357.2)
    (4, 712.8)
    (8, 376.7)
    (16, 215.2)
    (32, 125.2)
    (64, 65.6)
    (96, 70.5)
};

\legend{Silca-CKKS, SilcaZ-BGV}
\end{axis}
\end{tikzpicture}

}
\caption{Caching Overhead}
\label{fig:overhead}
\end{figure}

Figure~\ref{fig:overhead} illustrates the performance comparison of two encryption schemes, Silca-CKKS and SilcaZ-BGV, in terms of execution time (in milliseconds) with respect to the number of threads used. The x-axis represents the number of threads, while the y-axis represents the execution time.

The plot consists of two data series represented by blue and red markers. The blue markers correspond to the CKKS encryption scheme, while the red markers correspond to the BGV encryption scheme.

The data points on the plot indicate the average execution time observed for each configuration. As the number of threads increases from 1 to 96, the execution time for both encryption schemes generally decreases. However, there are certain points where the execution time slightly increases before decreasing again.

For the Silca-CKKS encryption scheme (represented by round markers), the execution time starts at 560.3 seconds for 1 thread, decreases to 29.5 seconds at 32 threads, and then slightly increases to 59.3 seconds at 96 threads.
We see that Silca-CKKS can reduce the caching overhead from about 10 minutes into less than one minute.

For the SilcaZ-BGV encryption scheme (represented by square markers), the execution time starts at 2,627.3 seconds for 1 thread, decreases to 65.6 seconds at 64 threads, and then slightly increases to 70.5 seconds at 96 threads.
Similarly, SilcaZ-BGV reduces the overhead from tens of minutes to only slightly more than one minute.



\subsection{Memory Footprint}

\begin{figure}
    \centering
  \resizebox{0.65\linewidth}{!}{ 
    
    \begin{tikzpicture}
    \begin{axis}[
        xlabel={Number of Threads},
        ylabel={Memory (GB)},
        xticklabels={1, 2, 4, 8, 16, 32, 64, 96},
        xtick={1, 2, 3, 4, 5, 6, 7, 8},
        ytick={0, 50, 100, 150, 200, 250},
        ymin=0,
        ymax=250,
        ymajorgrids=true,
        grid style=dashed,
        legend style={at={(0.02,0.98)}, anchor=north west},
        legend cell align=left,
        legend entries={Silca-CKKS, SilcaZ-BGV},
        legend style={draw=none},
    ]
    
    \addplot[color=green, mark=triangle*] coordinates {
        (1, 65)
        (2, 65)
        (3, 65)
        (4, 65)
        (5, 65)
        (6, 65)
        (7, 65)
        (8, 97)
    };
    
    \addplot[color=orange, mark=diamond*] coordinates {
        (1, 146)
        (2, 146)
        (3, 146)
        (4, 146)
        (5, 146)
        (6, 146)
        (7, 146)
        (8, 219)
    };
    
    \end{axis}
    \end{tikzpicture}
}
    \caption{Memory Footprint}
    \label{fig:memory}
\end{figure}
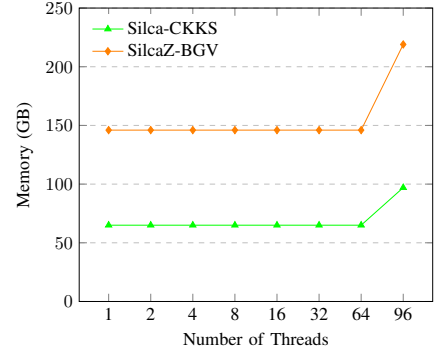

The line plot (Figure \ref{fig:memory}) illustrates the memory consumption of Streche-CKKS and Streche-BGV as the number of threads varies. The y-axis represents the memory usage in gigabytes (GB), and the x-axis corresponds to the number of threads employed.

For Silca-CKKS, the memory footprint remains constant at 65 GB across all thread configurations except for 96 threads, as depicted by the green line. 
We believe the case of 96 threads accounts for the hardware specification of our test bed:
the server is equipped with two physical 64-core CPUs;
therefore, when more than 64 threads are invoked,
additional overhead is possibly incurred between distinct physical CPUs.
Other than the 96-thread case,
our result indicates that Silca-CKKS exhibits consistent memory requirements regardless of the number of threads utilized. The uniform memory usage of Silca-CKKS suggests that it is well-suited for memory-constrained environments, as it maintains a predictable and manageable memory overhead.

SilcaZ-BGV, represented by the orange line, demonstrates a consistent memory consumption of 146 GB for thread counts ranging from 1 to 64. This suggests that SilcaZ-BGV also maintains stable memory requirements within this range. However, when employing 96 threads, the memory usage increases significantly to 219 GB. This substantial rise in memory footprint indicates a heightened demand for memory resources when utilizing a higher number of threads spanning across multiple CPU chips, 
as explained above.

\subsection{Silca with CKKS}

\begin{figure*}[!t]
  \centering
    \includegraphics[width=.8\linewidth]{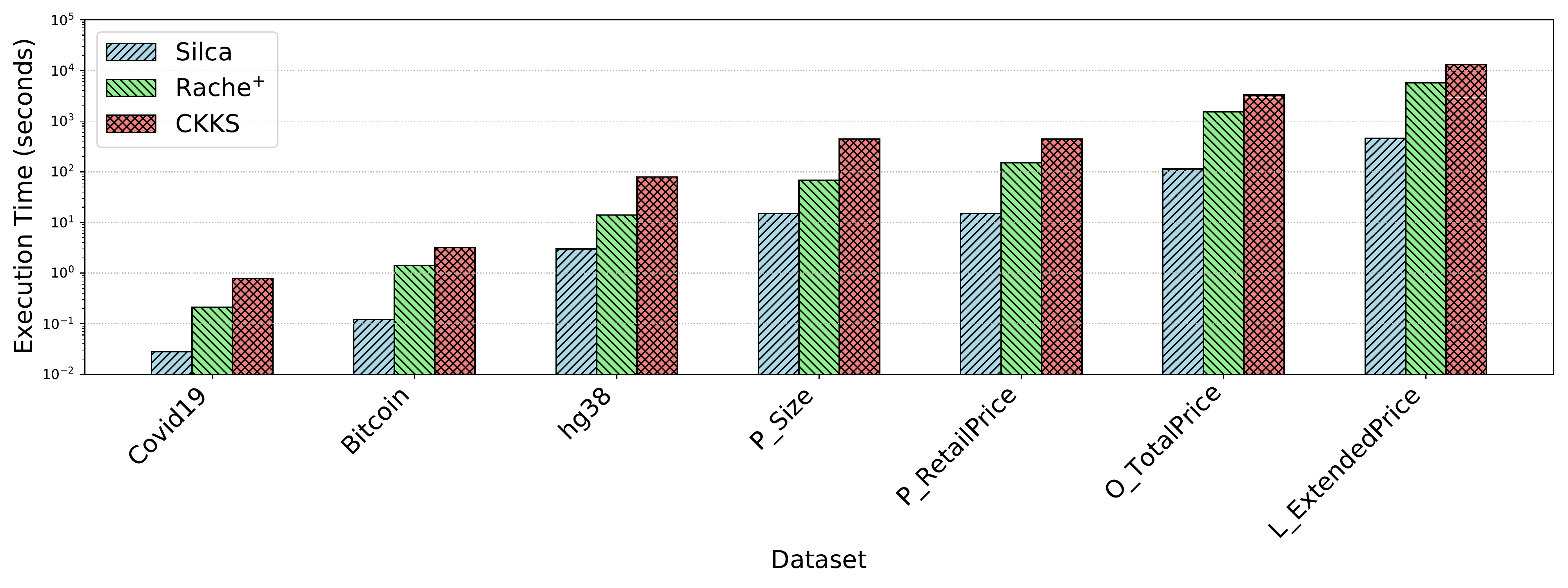}
  \caption{Comparing the encryption performance of different caching approaches for floating-point numbers.}
  \label{fig:ckks}
\end{figure*}

The presented figure, Figure \ref{fig:ckks}, illustrates a comparison of the encryption performance of floating-point numbers using three different caching approaches. The x-axis of the plot represents various benchmarks, including Covid19, Bitcoin, hg38, P\_Size, P\_RetailPrice, O\_TotalPrice, and L\_ExtendedPrice. These benchmarks serve as representative computational tasks or datasets for which the encryption performance is evaluated. 

The y-axis represents the execution time in seconds, displayed on a logarithmic scale. The logarithmic scale allows for a comprehensive representation of a wide range of execution times. The y-axis tick marks are positioned at values between 0.01 and 100,000.

The plot consists of three sets of horizontal bars, each corresponding to a specific caching approach. The length of each bar represents the execution time for a particular benchmark. The width of the bars is set to 2mm, ensuring a compact and visually appealing presentation of the data.

The first set of bars represents the performance of the Silca caching approach. These bars are filled with a pattern of vertical lines, distinguishing them from the other sets of bars. The lengths of the bars indicate the execution times for each benchmark, ranging from 0.028 seconds for Covid19 to 456 seconds for L\_ExtendedPrice.

The second set of bars represents the performance of the Rache$^+$ caching approach\footnote{The original Rache scheme does not support floating-point numbers; 
therefore, we extend the original Rache by splitting a floating-point number into a series of digits. We name this extended scheme as Rache$^{+}$.}. These bars are filled with a pattern of northwest lines, visually contrasting them from the other sets of bars. The lengths of these bars represent the execution times for each benchmark, ranging from 0.212 seconds for Covid19 to 5,754 seconds for L\_ExtendedPrice.

The third set of bars represents the performance of the CKKS caching approach. These bars are filled with a crosshatch pattern, distinguishing them from the other sets of bars. The lengths of these bars indicate the execution times for each benchmark, ranging from 0.783 seconds for Covid19 to 13,188 seconds for L\_ExtendedPrice.

In conclusion, Figure \ref{fig:ckks} offers a comprehensive visualization of the encryption performance of floating-point numbers using different caching approaches. The figure indicates that the Silca caching approach generally achieves faster execution times compared to Rache$^+$~\cite{otawose_sigmod23} and CKKS~\cite{ckks}.
The improvement is significant:
Silca is up to 10$\times$ faster than Rache$^+$ and orders of magnitude faster than CKKS.

\subsection{SilcaZ with BGV}

\begin{figure*}[t]
  \centering
    \includegraphics[width=.8\linewidth]{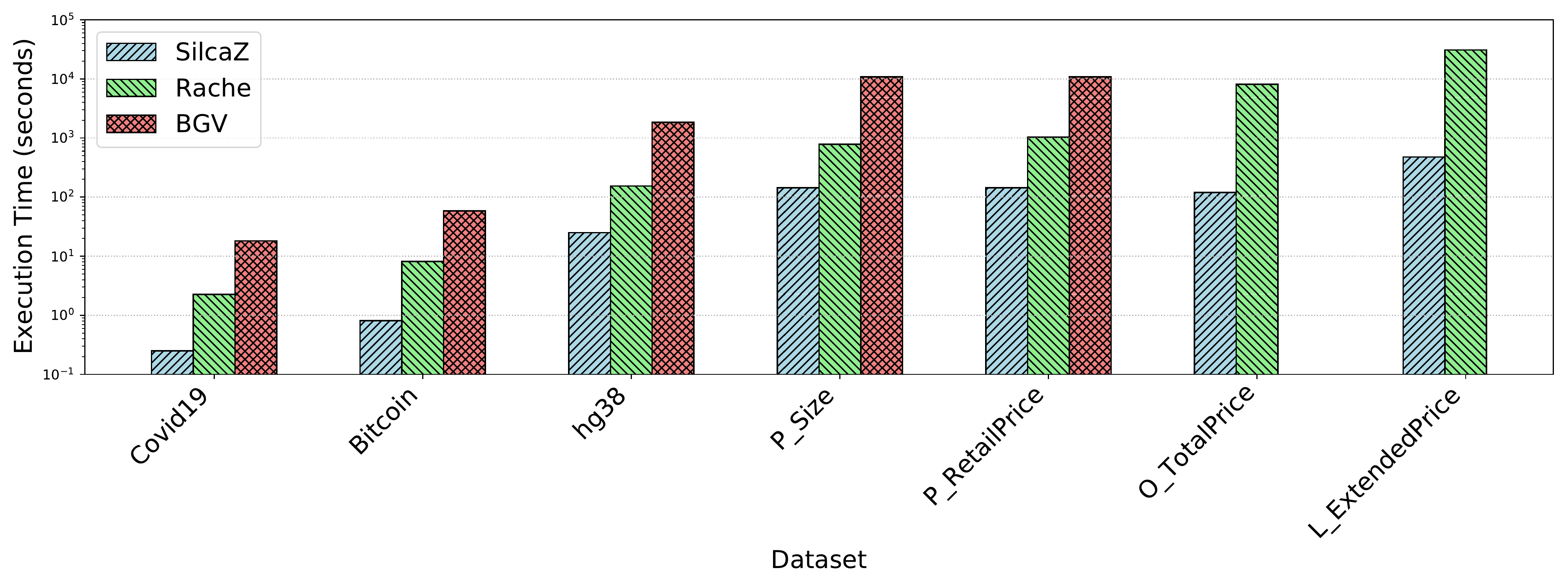}
  \caption{Comparing the encryption performance of integers using BGV~\cite{bgv}, Rache~\cite{otawose_sigmod23}, and SilcaZ (this work).}
  \label{fig:bgv}
\end{figure*}

The figure shown in Figure \ref{fig:bgv} presents a comparison of the encryption performance of integers using BGV, Rache, and SilcaZ.
The execution time (in seconds) for various benchmarks is plotted on the y-axis, while the corresponding benchmarks are labeled on the x-axis.

The x-axis of the plot represents different benchmarks, namely Covid19, Bitcoin, hg38, P\_Size, P\_RetailPrice, O\_TotalPrice, and L\_ExtendedPrice. These benchmarks represent different computational tasks or datasets for which the encryption performance is evaluated.

The y-axis represents the execution time in seconds. The scale is logarithmic. This scale allows for a better visualization of the data across a wide range of execution times.
The plot consists of three different sets of bars, each representing a different caching method. The bars are aligned to show the relative execution times for each benchmark. 

The first set of bars represents the performance of the SilcaZ method. The bars are filled with a pattern of vertical lines, distinguishing them from the other sets of bars. The lengths of the bars indicate the execution times for each benchmark, ranging from 0.25 seconds for Covid19 to 478.8 seconds for L\_ExtendedPrice.

The second set of bars represents the performance of the Rache method. These bars are filled with a pattern of northwest lines, providing a visual contrast with the other sets of bars. The lengths of these bars represent the execution times for each benchmark, ranging from 2.26 seconds for Covid19 to 30,960 seconds for L\_ExtendedPrice.

The third set of bars represents the performance of the BGV method. These bars are filled with a crosshatch pattern, differentiating them from the other sets of bars. However, for the benchmarks P\_RetailPrice and O\_TotalPrice, the execution times are over 24 hours and therefore are not available, resulting in missing bars.

This figure allows for a visual comparison of the encryption performance of integers using different caching methods. It shows that the SilcaZ method generally exhibits orders of magnitude faster execution times compared to the BGV and the Rache methods across most benchmarks.

\section{Conclustion and Related Work}
\label{sec:conc}

In this work, we have made significant contributions to improving the performance of homomorphic encryption for outsourced databases in cloud computing. Our proposed optimizations, Silca and SilcaZ, address the computational overhead and efficiency challenges associated with homomorphic encryption techniques.

Silca leverages advanced caching strategies and indexing techniques to reduce the computational overhead of encryption operations. By intelligently managing encrypted data and optimizing access patterns, Silca minimizes latency and improves the overall performance of computations on encrypted data. Building upon Silca, SilcaZ takes advantage of efficient modular arithmetic operations and probabilistic data structures to enhance encryption performance in query-intensive outsourced databases. SilcaZ introduces efficient query processing techniques that minimize communication and computational costs while ensuring data privacy and integrity.

In addition to the practical implementations of Silca and SilcaZ as loadable functions in MySQL, we have provided a theoretical contribution by proving the semantic security of both optimizations. We demonstrated that breaking the security of Silca or SilcaZ would lead to breaking the security of the underlying homomorphic encryption scheme. This theoretical analysis provides a strong guarantee of the security properties of our proposed optimizations.

Through comprehensive experiments and evaluations using seven real-world datasets, we have also demonstrated the effectiveness and efficiency of Silca and SilcaZ in practice. Our results have shown significant performance improvements in terms of execution time and resource utilization compared to existing homomorphic encryption schemes. These experimental findings validate the practical viability and impact of our proposed optimizations.

By bridging the gap between theoretical advancements and practical applications, our research contributes to the field of outsourced databases in cloud computing. The system implementation of Silca and SilcaZ as loadable functions in MySQL further reinforces their practical applicability and provides a user-friendly integration into existing database management systems. This system implementation allows users to seamlessly incorporate Silca and SilcaZ into their database workflows, enabling secure and efficient outsourcing of sensitive data in cloud computing environments.


While our research has yielded promising results, there are several avenues for future work and further exploration:

\paragraph{Enhanced Security Analysis} While we have proven the semantic security of Silca and SilcaZ, conducting a more comprehensive security analysis that considers other potential attacks and vulnerabilities would be valuable. Exploring the resistance against advanced cryptanalytic techniques and conducting formal proofs can further strengthen the security guarantees of our proposals.

\paragraph{Optimization Techniques} Continuously improving the performance of Silca and SilcaZ is a critical aspect of future work. Exploring additional optimization techniques, such as parallelization, distributed computing, or hardware acceleration, can further enhance the efficiency and scalability of homomorphic encryption for outsourced databases.

\paragraph{Standardization and Integration} Promoting the adoption and integration of Silca and SilcaZ into existing standards and frameworks is essential for practical deployment. Collaborating with standardization bodies, database management system vendors, and cloud service providers can facilitate the integration process and ensure interoperability across different platforms.

\paragraph{Real-World Deployment and Case Studies} Conducting real-world deployments and case studies of Silca and SilcaZ in various application domains can provide valuable insights into their practical usability, performance, and impact. Evaluating the optimizations in different scenarios and assessing their performance in large-scale production environments will further validate their effectiveness and guide their usage in real-world settings.

By pursuing these future directions, we can continue to advance the field of secure cloud computing, making it more efficient, secure, and applicable to real-world scenarios involving outsourced database systems. The system implementation of Silca and SilcaZ in MySQL loadable functions serves as a foundation for practical deployment and encourages further exploration of their potential applications and benefits.

\bibliographystyle{IEEEtran}
\bibliography{ref_new}

\end{document}